\documentclass[12pt,preprint]{aastex}

\newcommand{\rrab}{RR{\sl ab}}
\newcommand{\rrc}{RR{\sl c}}

\slugcomment{To appear in AJ}
\shorttitle{RR Lyrae stars in the Hydra II galaxy}
\shortauthors{A. K. Vivas et al.}

\begin{document}

\title{Variable stars in the field of the Hydra II ultra-faint dwarf galaxy}

\author{A. Katherina Vivas$^{1}$, Knut Olsen$^{2}$, Robert Blum$^{2}$,
David L. Nidever$^{3,4,5}$
Alistair R. Walker$^{1}$, Nicolas F. Martin$^{6,7}$, Gurtina Besla$^{5}$, Carme Gallart$^{8,9}$, 
Roeland P. van der Marel$^{10}$, Steven R. Majewski$^{11}$, Catherine C. Kaleida$^{1}$,
Ricardo R. Mu\~noz$^{12}$, Abhijit Saha$^{2}$, Blair C. Conn$^{13}$, Shoko Jin$^{14}$
}

\email{kvivas@ctio.noao.edu}

\altaffiltext{1}{Cerro Tololo Inter-American Observatory, National Optical Astronomy Observatory, Casilla 603, La Serena, Chile}
\altaffiltext{2}{National Optical Astronomy Observatory, 950 N Cherry Ave, Tucson, AZ 85719, USA}
\altaffiltext{3}{Department of Astronomy, University of Michigan, 1085 S. University Ave., Ann Arbor, MI 48109-1107, USA}
\altaffiltext{4}{Large Synoptic Survey Telescope, 950 North Cherry Ave, Tucson, AZ 85719}
\altaffiltext{5}{Steward Observatory, 933 North Cherry Ave, Tucson, AZ 85719} 
\altaffiltext{6}{Observatoire astronomique de Strasbourg, Universit\'e de Strasbourg, CNRS, UMR 7550, 11 rue de l'Universit\'e, F-67000 Strasbourg, France}
\altaffiltext{7}{Max-Planck-Institut f\"ur Astronomie, K\"onigstuhl 17, D-69117 Heidelberg, Germany}
\altaffiltext{8}{Instituto de Astrof\'{i}sica de Canarias, La Laguna, Tenerife, Spain}
\altaffiltext{9}{Departamento de Astrof\'{i}sica, Universidad de La Laguna, Tenerife, Spain}
\altaffiltext{10}{Space Telescope Science Institute, 3700 San Martin Drive, Baltimore, MD 21218}
\altaffiltext{11}{Department of Astronomy, University of Virginia, Charlottesville, VA 22904, USA}
\altaffiltext{12}{Departamento de Astronom\'ia, Universidad de Chile, Camino del Observatorio 1515, Las Condes, Santiago, Chile}
\altaffiltext{13}{Gemini Observatory, Recinto AURA, Colina El Pino s/n, La Serena, Chile.}
\altaffiltext{14}{Kapteyn Astronomical Institute, University of Groningen, P.O. Box 800, 9700 AV Groningen, The Netherlands}

\begin{abstract}

We report the discovery of one RR Lyrae star in the ultra--faint satellite galaxy Hydra II based on time series 
photometry in the $g$, $r$ and $i$ bands obtained with the Dark Energy Camera at Cerro Tololo Inter-American Observatory, Chile. 
The association of the RR Lyrae star discovered here with Hydra II is clear because is located at $42\arcsec$ from the center of the dwarf, 
well within its half-light radius of  $102\arcsec$. 
The RR Lyrae star has a mean magnitude of  $i = 21.30\pm 0.04$ which is too faint to be a field halo star. This magnitude translates to
a heliocentric distance of $151\pm 8$ kpc for Hydra II; this value is $\sim 13\%$  larger than the estimate from the discovery paper based on the average magnitude of several blue horizontal branch star candidates. 
The new distance implies a slightly larger half-light radius of $76^{+12}_{-10}$ pc and a brighter absolute magnitude of $M_V = -5.1 \pm  0.3$, which keeps this object within the realm of the dwarf galaxies. 
A comparison with other RR Lyrae stars  in ultra--faint systems indicates similar pulsational properties among them, which are different to those found among halo field stars and those in the largest of the Milky Way satellites. We also report the discovery of 31 additional short period variables in the field of view (RR Lyrae, SX Phe,
eclipsing binaries, and a likely anomalous cepheid) which are likely not related with Hydra II.

\end{abstract}

\keywords{galaxies: dwarf; galaxies: individual (Hydra II); Local Group; stars: variables: RR Lyrae;
stars: variables: general}

\section{Introduction}

Hydra II is one of the new faint satellite galaxies of the Milky Way recently discovered in the Southern 
hemisphere with the 
Dark Energy Camera \citep{bechtol15,koposov15,martin15,kim15a,kim15b,drlica15}. The galaxy was
found by searching for overdensities in data from the Survey of the Magellanic Stellar History 
\citep[SMASH,][]{martin15}.
The discovery of new low-luminosity satellites is of tremendous importance 
for reconciling observations and LCDM predictions for the number and distribution of 
satellites around the Milky Way \citep[the missing satellite problem,][]{klypin99,moore99}.
Several of the new galaxies, including Hydra II, 
have the added aspect of a possible 
association with the Magellanic Clouds \citep{martin15,bechtol15}. 
This is an interesting scenario (that needs further investigation)
because cosmological numerical simulations predict that cases of ``satellites of satellites" are common
at the time of infall \citep{donghia08,sales11,deason15,wheeler15}. 
Confirmation and full characterization of the new discoveries (e.g., establishing them as satellite galaxies rather than globular clusters) are needed in order to test 
these theoretical predictions and to understand the role of dwarf systems in the formation of Milky Way--type galaxies.
Key to such characterization is accurate distance measurements toward the new galaxies. 

\citet{kirby15} recently obtained spectroscopy for several stars in Hydra II. 
Although they could not resolve 
the velocity dispersion of the system (the main diagnostic for confirming it as a galaxy), the large 
metallicity dispersion they measured ($\sigma$[Fe/H]$=0.4$),
together with Hydra II's relatively large size \citep[68 pc,][but see revised value below]{martin15} makes it more likely to be a dwarf 
galaxy rather than a globular cluster.

Variable stars have a long tradition of being an excellent tool for studying the content and structure 
of stellar systems.  The presence of RR Lyrae stars is an unequivocal sign of an old stellar population 
($> 10$ Gyr) since these stars are evolved sub--solar--mass stars ($\sim 0.7 M_{\odot}$) burning 
helium in their cores \citep{smith95}. 
All of the dwarf galaxies for which time series data are available have shown the 
presence of at least one RR Lyrae star \citep[see compilations in][]{vivas06,boettcher13,baker15}, 
which confirms that
all dwarf galaxies do have a population of old stars;
this includes Segue 1 ($M_V=-1.5$), the
lowest luminosity galaxy with known RR Lyrae stars \citep{simon11}. 
The detection of RR Lyrae stars in low--luminosity systems is also important for other reasons. 
RR Lyrae stars are standard candles
and thus provide an alternative method for determining the distances to galaxies. This is
particularly critical in these distant and low--luminosity systems lacking young stellar populations
(i.e., systems that cannot host Cepheids, which could also be used as distance indicators). 
In addition, it is hard to fit 
isochrones to low luminosity systems because there are very few stars in the upper part of the color--magnitude
diagram (CMD), the contamination by field 
stars may be important due to the low density of stars in the system, and the main sequence turn--off 
may not be available in some systems due to their large distances. In the case of Hydra II, 
for example, the 
distance was estimated by \citet{martin15} by identifying 12 probable horizontal branch (HB) stars in the 
galaxy and using the HB
absolute magnitude -- color relationship of \citet{deason11}. If Hydra II contains RR Lyrae stars, the 
distance could be accurately calculated and important parameters such as the physical half--light radius
and absolute magnitude refined. 
In addition, the comparison of the pulsational properties of RR Lyrae in the halo field and in the MIlky Way satellites allows us to study the origin of the field population of RR Lyrae stars, 
and thus the formation of the Galactic halo. Based on their pulsational properties, it has been argued that most of the Galactic field RR Lyrae stars may 
have come from a few mergers with Sgr--like galaxies \citep{zinn14,fiorentino15}. 
The ultra--faint satellites may be an important source to the long-period tail of the period distribution of field RR Lyrae stars
in the halo, an idea that can be confirmed as more data on low--luminosity galaxies are accumulated.
  
On the other hand, the presence of other types of variable stars are important for studying the stellar population in a 
galaxy. Anomalous cepheids (which are brighter than RR Lyrae stars) are usually
interpreted as belonging to metal--poor, intermediate--age population \citep[e.g.,][]{fiorentino12a,fiorentino12b,ripepi14}. 
Their progenitors are likely stars with $\sim 1-2 M_\odot$ in the He-burning phase, although it has been suggested they can also
be the result of binary interaction.
The ultra--faint galaxy Leo T, for example, 
contains 11 such stars \citep{clementini12}, suggesting a very complex star formation
history for this galaxy. Fainter pulsating stars, also known as dwarf cepheids, may
be present as well \citep[e.g.][]{breger00,mcnamara11,vivas13}. Hydra II is suspected to have a blue straggler population \citep{martin15} and some
of them may be variable stars of the SX Phe type \citep[see for example][]{vivas13}.

In this work we explore the (short--period, $<1$ day) variable population of the ultra--faint dwarf Hydra II and
its surroundings. 

\section{Observations}

\begin{figure}[ht]
\epsscale{0.7}
\plotone{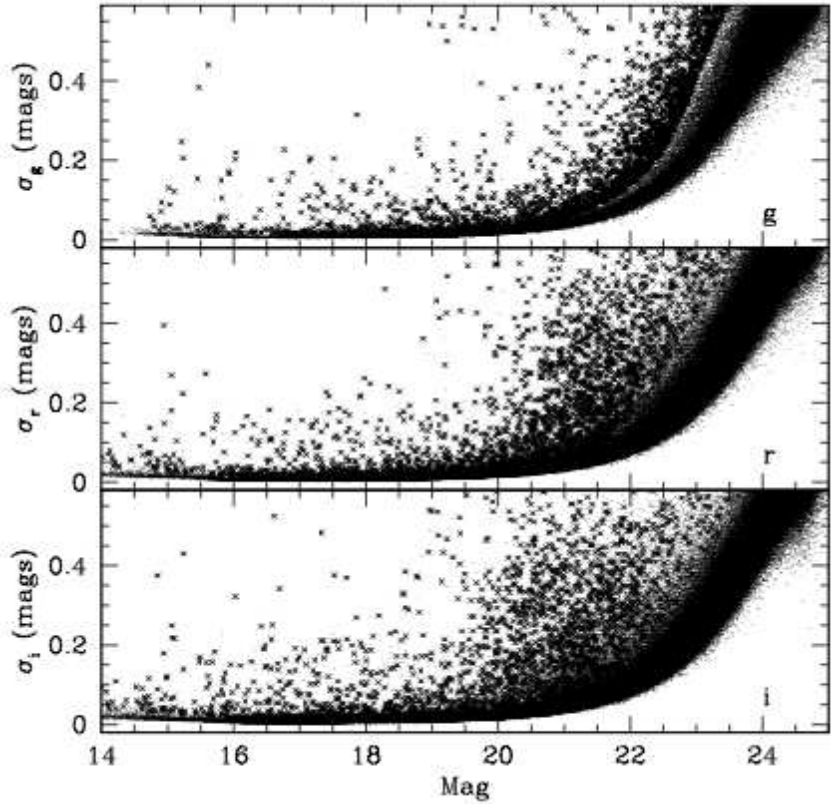}
\caption{The standard deviation in magnitude for all the stars in the field as a 
function of their mean magnitude. The main locus is formed by non--variable stars. Stars above
these loci are candidates variable stars. Crosses represent stars whose variations in magnitude in all three bands ($g$, $r$ and $i$) are significantly larger than the expected error at their mean magnitude. }
\label{fig-error}
\end{figure}

Observations were carried out during 2015 March 30--31 with the Dark Energy Camera 
\citep[DECam,][]{flaugher15} at the
Blanco telescope at the Cerro Tololo Inter-American Observatory (CTIO), Chile.
Consecutive observations of Hydra II ($\alpha=12^\mathrm{h}21^\mathrm{m}42.1^\mathrm{s}, \delta=-31\degr59\arcmin07\arcsec$) 
were taken during $\sim 8$ hours both nights, alternating the $g$, $r$ and $i$ filters.
Because these nights were close to full moon, exposure times were kept relatively short (150s) in order 
to keep the background flux to manageable levels. Even in bright time, this exposure time is good enough 
to obtain an excellent S/N at the level of the HB of Hydra II ($g\sim21.5$).  The typical time separation between observations in the
same band was $\sim 8.5$ min. 
During the 
short period with no moon on these nights we took several continuous, deeper exposures (360s) in $g$ and $r$. 
In total we obtained 111, 112 
and 102 observations in $g$, $r$ and $i$, respectively.
The seeing 
was stable during the whole observing run at around $\sim 1.0$ arcsec. 
The galaxy was centered on DECam chip S4 at the center of the field of view.
The entire galaxy \citep[$r_h=1\farcm 7$,][]{martin15} fits well within a single CCD, which has a FOV 
of $9\arcmin \times 18\arcmin$. However, we searched for variables in the full DECam FOV ($2\fdg 2$  diameter)
to explore for a possible extra-tidal population of variable stars.

Images were processed by the DECam Community Pipeline \citep{valdes14} which performs bias 
subtraction, flatfielding and astrometric calibration, among other 
steps\footnote{\url{http://ast.noao.edu/sites/default/files/NOAO\_DHB\_v2.2.pdf}}.
PSF photometry was made using the PHOTRED\footnote{\url{https://github.com/dnidever/PHOTRED}} 
pipeline \citep{nidever11} which is based on DAOPHOT - ALLSTAR
routines \citep{stetson87,stetson94}. To isolate stars from extended objects,
the final catalog was cleaned by cutting out objects having
(Sextractor) values of $\vert{\rm sharp}\vert>1.0$, ${\rm chi}>3.0$ and ${\rm prob}<0.8$. 

Then, using as reference the image with the lowest airmass observation in each band 
we performed relative 
photometry by calculating zero-point differences between each individual observation and the reference
image. Because both nights were photometric, the differences in zero point among the different observations
were due only to variations in atmospheric extinction due to the changing airmass of the field
during each night. Once the zero point differences were applied to the data, mean magnitudes 
for each star were calculated. The resulting catalog was calibrated using
these (instrumental) mean magnitudes and comparing them with the photometry measured in the
discovery paper \citep{martin15}. Both zero points and color terms were included in the transformation equations.
The final ensemble contains calibrated $g$, $r$ and $i$ time series for 
277,579 point sources. 

Figure~\ref{fig-error} shows the standard deviation ($\sigma_{\rm star}$) 
of the magnitudes of each star as a function of
their mean magnitude. The main locus observed in each of these figures corresponds to non-variable stars and
defines the photometric errors of our observations as a function of magnitude. To characterize the loci
we binned the data in 0.2 mag bins and calculated the (sigma-clipped) mean error ($\sigma(m)$) 
and standard deviation (${\rm std}(m)$) in each bin centered on magnitude $m$.  
The mean photometric errors of the sample are $<0.01$ mags for $g,r,i \lesssim 19.5$ and 
increase to 0.1 mags around $m=22$ in each band. The expected error at the level of
the HB of Hydra II is $\sim 0.05$ mags.
We consider stars as variables if they are located significantly above the locus of the normal stars ($\sigma_{\rm star} > \sigma(m) + 
3 \times {\rm std}(m)$). About $\sim 1\%$ (3,718 stars) of all stars in the field show variability simultaneously
in all three bands. These variable stars are indicated with crosses in Figure~\ref{fig-error}.

\begin{figure}[ht]
\epsscale{0.5}
\plotone{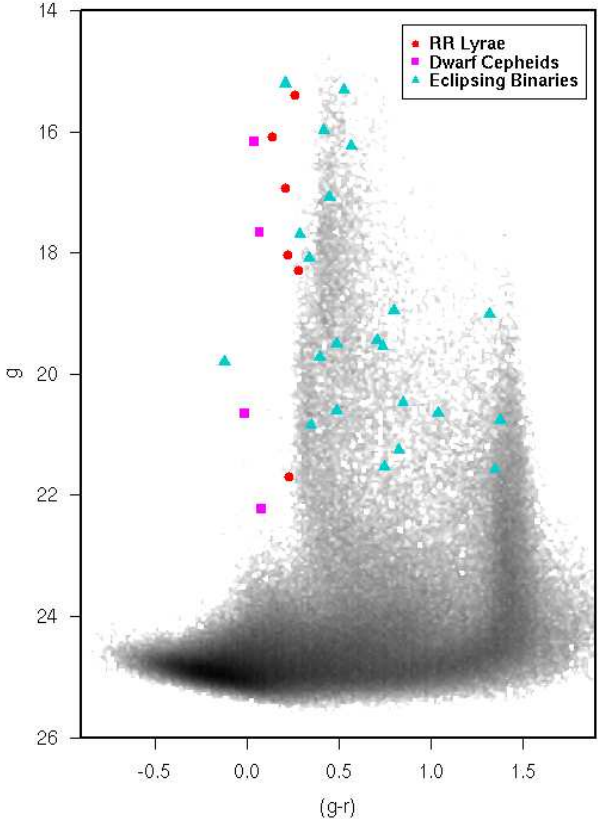}
\caption{Color-Magnitude diagram (log density plot) of the full DECam field of view. At this scale, the features
of the Hydra II galaxy are not noticeable. The large symbols indicate the location of the variable stars
found in the field.}
\label{fig-CMDall}
\end{figure}

\section{Periodic Variable Stars}

We proceed to identify periodic variable stars in the dataset by using the well-known phase-dispersion minimization 
method first described by \citet[][LK]{lafler65}. To make full use of all
the available multi-band data, we combined the results of the LK method in each individual band to derive an enhanced
string length parameter \citep[see for example][]{watkins09,mateu12,vivas13}:

\begin{equation}
\Theta = \frac {N_g\Theta_g + N_r\Theta_r + N_i\Theta_i} {N_g+N_r+N_i}
\end{equation}

\noindent
where we weight by the number of observations
in each band, $N_\lambda$.

In the original LK method,
the correct period is obtained by finding the minimum value of the string length parameter ($\Theta_\lambda$) for a 
range of trial periods in a particular band.
The parameter $\Theta_\lambda$ is the mean squared difference between magnitudes ($m_i$) at consecutive phases for a given trial period:

\begin{equation}
\Theta_\lambda = \frac{\sum_i  (m_i-m_{i+1})^2}{\sum_i (m_i-\overline{m})^2}
\end{equation}

Because our observations contain data from only two nights, the period search was restricted to values 
less than one day. Our data are not sensitive to long-period variables. We applied the method in two 
period ranges, one optimized for RR Lyrae stars and anomalous cepheids (0.15 to 1.0 days), 
and the other one aimed at finding  dwarf cepheid stars (0.01 to 0.15 days). We considered several
minima of $\Theta$ during our search to allow for possible spurious periods due to harmonics or aliases.

After visually examining potential periodic stars given by this method, we finally selected and classified 
32 periodic variables in the field. Most of the periodic stars are eclipsing binaries, but we also
found six RR Lyrae stars and four dwarf cepheids (one of them is dubious and may be an anomalous cepheid).
The rest of the $\sim 3,000$ variables are probably due to noise and artifacts (half of them are fainter than $g=22.2$), 
non-periodic variables, or stars with periods $>1$ day.
Table~\ref{tab-var} contains coordinates, number of observations in each band, mean magnitudes,
amplitude, period and classification for all the periodic variable stars (see Sections 3.1 to 3.3).
We refer the reader to \citet{drake14}
for a general description of the observational properties of the different types of variables. 

We searched the International Variable Star Index (VSX)\footnote{http://www.aavso.org/vsx/index.php}
and found that only one star out of the 32 had been previously classified as a periodic variable.
This star is 48.10426, which has been classified as a {\rrab } star in the 
Catalina Real-Time Transient Survey \citep[CRTS]{torrealba15}, under ID SSS\_J122306.8-322119, 
with a period similar to that we have found in this work. 
The remaining 31 stars should be considered
as new discoveries. Star 48.10426 is the brightest of the RR Lyrae stars in our sample. Fainter stars may have 
been missed by the CRTS since its completeness starts to decrease at magnitude $\sim 16$ 
\citep{torrealba15}. 

A color-magnitude diagram (CMD) of the full DECam field is shown in Figure~\ref{fig-CMDall}, together with the periodic
variable stars found within it. This CMD was obtained by averaging the multiple individual observations of each star.
As expected, RR Lyrae stars and dwarf cepheids (red circles and magenta squares) 
are located in a narrow range of color in this diagram ($0.00<(g-r)<0.28$, the instability strip), while eclipsing binaries are 
widely spread in color. 

\begin{figure}[ht]
\epsscale{0.7}
\plotone{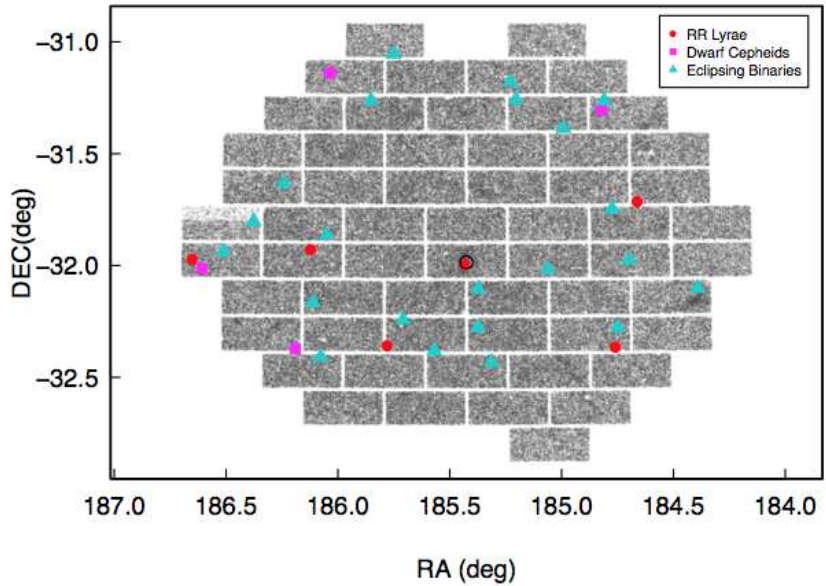}
\caption{Spatial distribution of the 277,579 point-like sources in the DECam field. 
Hydra II is located in one of the central chips and
is marked with a black circle, whose size reflects the half-light radius ($1\farcm 7$)
of the galaxy. One of the amplifiers in chip S7 (central, leftmost) has unstable gain, which results in no reliable
photometry in the upper part of the chip. 
Variable stars are indicated with large color symbols. One RR Lyrae star lies within the half-light radius of Hydra II.}
\label{fig-sky}
\end{figure}

Figure~\ref{fig-sky} shows the location of the variable stars in the field.\footnote{Notice there is no data in chip N31, which is located on the east side of the southernmost row of CCDs.
This was due to an error in the photometry pipeline, which skipped this CCD. Since this chip is far enough from Hydra II, its inclusion would not have changed the results discussed in this paper.}  
The circle in one of the central CCDs
indicates the location and half-light radius of Hydra II. Only one of the variable stars, an RR Lyrae star, lies within the
half-light radius of the galaxy\footnote{This result was confirmed by an alternative method for searching variable stars, described in \citet{bernard09}.}. 
Thus, the rest of the variable stars are likely to be Milky Way field stars.

\begin{deluxetable}{lcccccccccccccc}
\tabletypesize{\tiny}
\rotate
\tablecolumns{14}
\tablewidth{0pc}
\tablecaption{Variable Stars in the Hydra II Field}
\tablehead{
ID   &   RA(2000.0) &  DEC(2000.0) &  $N_g$ & g & $\Delta g$ & $N_r$ & r & $\Delta r$ & $N_i$ & i & Amp i & Period (d) & Type \\
}
\startdata
  3.4557 & 12:23:00.33 & -31:03:15.0 &  56 & 19.55 & 0.77 &  57 & 18.81 & 0.74 &  51 & 18.45 & 0.75 & 0.233 &   EW \\ 
  5.5304 & 12:20:55.22 & -31:10:48.3 & 109 & 21.54 & 0.72 & 112 & 20.79 & 0.54 & 102 & 20.38 & 0.45 & 0.228 &   EB \\ 
 7.16439 & 12:24:08.42 & -31:08:23.6 &  81 & 16.17 & 0.73 &  76 & 16.13 & 0.53 &  69 & 16.09 & 0.39 & 0.0589 &   DC \\ 
     8.9 & 12:19:14.43 & -31:15:46.2 & 107 & 19.79 & 0.26 & 108 & 19.91 & 0.25 & 102 & 19.92 & 0.22 & 0.340 &   EW \\ 
  8.9705 & 12:19:17.35 & -31:18:18.2 & 111 & 20.66 & 0.63 & 112 & 20.67 & 0.50 & 102 & 20.65 & 0.38 & 0.0447 &   DC \\ 
  9.3499 & 12:19:58.30 & -31:23:07.6 &  56 & 16.24 & 0.72 &  57 & 15.67 & 0.71 &  50 & 15.42 & 0.69 & 0.270 &   EW \\ 
  9.8816 & 12:20:48.77 & -31:15:28.2 & 107 & 20.60 & 0.19 & 108 & 20.11 & 0.16 & 102 & 19.87 & 0.13 & 0.120 &   EW \\ 
 11.8451 & 12:23:24.56 & -31:15:44.3 & 107 & 18.96 & 0.12 & 108 & 18.16 & 0.09 & 102 & 17.80 & 0.06 & 0.629 &   EW \\ 
19.11108 & 12:18:38.43 & -31:42:51.1 &  56 & 16.93 & 0.78 &  57 & 16.72 & 0.53 &  51 & 16.57 & 0.41 & 0.599 & RRab \\ 
 24.1361 & 12:24:58.18 & -31:38:02.2 &  76 & 17.08 & 0.55 &  93 & 16.63 & 0.52 &  90 & 16.40 & 0.49 & 0.360 &   EW \\ 
 26.8646 & 12:19:05.84 & -31:44:49.1 & 107 & 18.09 & 0.13 & 108 & 17.75 & 0.11 & 102 & 17.57 & 0.11 & 0.173 &   EW \\ 
 30.2585 & 12:24:12.26 & -31:51:47.9 & 111 & 15.97 & 0.16 & 111 & 15.55 & 0.19 &  99 & 15.34 & 0.17 & 0.171 &   EW \\ 
 31.1203 & 12:25:30.57 & -31:48:10.8 & 106 & 21.58 & 1.07 & 107 & 20.23 & 0.60 &  98 & 19.38 & 0.50 & 0.290 &   EA \\ 
33.15473 & 12:18:47.89 & -31:58:31.4 & 111 & 15.20 & 0.30 & 104 & 14.99 & 0.33 &  97 & 14.88 & 0.29 & 0.665 &   EW \\ 
 34.6157 & 12:20:15.00 & -32:00:57.5 & 111 & 19.45 & 0.11 & 112 & 18.74 & 0.10 & 102 & 18.40 & 0.08 & 0.575 &   EW \\ 
 35.6516 & 12:21:43.51 & -31:59:42.8 & 111 & 21.70 & 0.68 & 112 & 21.47 & 0.48 & 102 & 21.30 & 0.38 & 0.645 & RRab \\ 
 37.4763 & 12:24:28.77 & -31:55:27.0 & 111 & 18.28 & 0.47 & 112 & 18.00 & 0.33 & 102 & 17.84 & 0.28 & 0.658 & RRab \\ 
 38.5537 & 12:26:02.85 & -31:56:20.6 & 111 & 20.46 & 0.14 & 112 & 19.61 & 0.13 & 102 & 18.67 & 0.09 & 0.145 &   EW \\ 
 38.8646 & 12:26:26.01 & -32:00:43.8 & 111 & 17.66 & 0.80 & 112 & 17.59 & 0.49 & 102 & 17.53 & 0.35 & 0.275 &   DC \\ 
 38.9496 & 12:26:35.89 & -31:58:13.7 & 111 & 16.08 & 0.48 & 112 & 15.94 & 0.33 & 102 & 15.85 & 0.25 & 0.336 &  RRc \\ 
 39.2293 & 12:17:34.22 & -32:06:04.0 &  90 & 17.70 & 0.27 &  83 & 17.41 & 0.24 &  65 & 17.26 & 0.24 & 0.375 &   EW \\ 
41.11296 & 12:21:29.40 & -32:06:15.2 & 111 & 15.30 & 0.52 &  83 & 14.77 & 0.43 &  67 & 14.52 & 0.38 & 0.290 &   EW \\ 
43.11303 & 12:24:26.73 & -32:09:55.7 & 108 & 19.02 & 0.37 & 108 & 17.70 & 0.29 & 100 & 16.93 & 0.30 & 0.695 &   EA \\ 
 46.2031 & 12:18:59.71 & -32:16:39.4 & 110 & 20.77 & 0.33 & 112 & 19.39 & 0.27 & 101 & 18.56 & 0.20 & 0.250 &   EB \\ 
 46.2447 & 12:19:02.16 & -32:21:55.5 &  69 & 18.03 & 1.12 &  66 & 17.81 & 0.88 &  69 & 17.60 & 0.68 & 0.535 & RRab \\ 
 47.9286 & 12:21:29.88 & -32:16:38.3 & 111 & 19.50 & 0.63 & 112 & 19.01 & 0.58 & 102 & 18.77 & 0.56 & 0.355 &   EW \\ 
48.10426 & 12:23:07.03 & -32:21:15.8 & 106 & 15.40 & 0.61 & 100 & 15.14 & 0.41 &  85 & 14.99 & 0.34 & 0.604 & RRab \\ 
 48.4574 & 12:22:16.57 & -32:22:54.2 &  55 & 19.72 & 0.11 &  57 & 19.32 & 0.08 &  51 & 19.10 & 0.09 & 0.139 &   EW \\ 
 48.8938 & 12:22:50.08 & -32:14:34.1 & 107 & 21.26 & 0.81 & 108 & 20.43 & 0.69 & 102 & 20.06 & 0.66 & 0.213 &   EW \\ 
  50.108 & 12:24:45.82 & -32:22:19.0 &  56 & 22.23 & 1.04 &  57 & 22.15 & 0.81 &  51 & 22.08 & 0.60 & 0.0611 &   DC \\ 
 53.1299 & 12:21:15.21 & -32:25:55.8 & 108 & 20.64 & 0.47 & 107 & 19.60 & 0.41 & 101 & 19.07 & 0.41 & 0.295 &   EB \\ 
 55.3234 & 12:24:18.22 & -32:24:29.8 & 105 & 20.84 & 0.23 & 108 & 20.49 & 0.19 & 102 & 20.29 & 0.19 & 0.214 &   EW \\
\enddata
\label{tab-var}
\tablecomments{ID nomenclature stands by chip\_number.star\_number (for example, 35.6516 is star
number 6516 in CCD 35); $N_g$, $N_r$ and $N_i$ are the number of observations for a given star
in each band. The different types of stars are: RRab and RRc = RR Lyrae stars of the type ab and c;
DC = dwarf cepheids; EW = eclipsing contact binaries; EB = semi--detached eclipsing binaries; EA =
detached eclipsing binaries.}
\end{deluxetable}

\begin{deluxetable}{lcccccccc}
\tabletypesize{\footnotesize}
\tablecolumns{9}
\tablewidth{0pc}
\tablecaption{Distance to RR Lyrae Stars in the Hydra II Field}
\tablehead{
ID   &   E(B-V) & $A_i$ & $A_V$ & $i_0$ & $V_0$ & $\mu_0$ (i) & Distance & Distance \\
      &              &           &            &           &           &   & from $M_i$ (kpc) & from $M_V$(kpc) \\
}
\startdata
19.11108 & 0.053 & 0.10 & 0.16 & 16.47 & 16.64 & 16.03$\pm$	0.08 &16.1$\pm$0.6 & 16.5$\pm$1.2 \\ 
{\bf 35.6516}   & {\bf 0.053} & {\bf 0.10} & {\bf 0.16}  & {\bf 21.19} & {\bf 21.40} & 
{\bf 20.89$\pm$0.11} & {\bf 151$\pm$8} & {\bf 154$\pm$12}  \\
37.4763   & 0.061 & 0.12 & 0.19 & 17.72 & 17.93 & 17.33$\pm$0.08 & 29$\pm$1 & 30$\pm$2 \\
38.9496   & 0.075 & 0.15 & 0.23 & 15.71 & 15.77 & 15.01$\pm$0.08 & 10.9$\pm$0.4 & 11.0$\pm$0.8 \\
46.2447   & 0.072 & 0.14 & 0.23 & 17.46 & 17.67 & 16.98$\pm$0.09 & 25$\pm$1 & 27$\pm$2 \\
48.10426 & 0.067 & 0.13 & 0.21 & 14.86 & 15.04 & 14.43$\pm$0.09 & 7.7$\pm$0.3 & 7.9$\pm$0.6 \\
\enddata
\label{tab-RR}
\end{deluxetable}

\subsection{RR Lyrae stars}

\begin{figure}[ht]
\center
\includegraphics[scale=0.5]{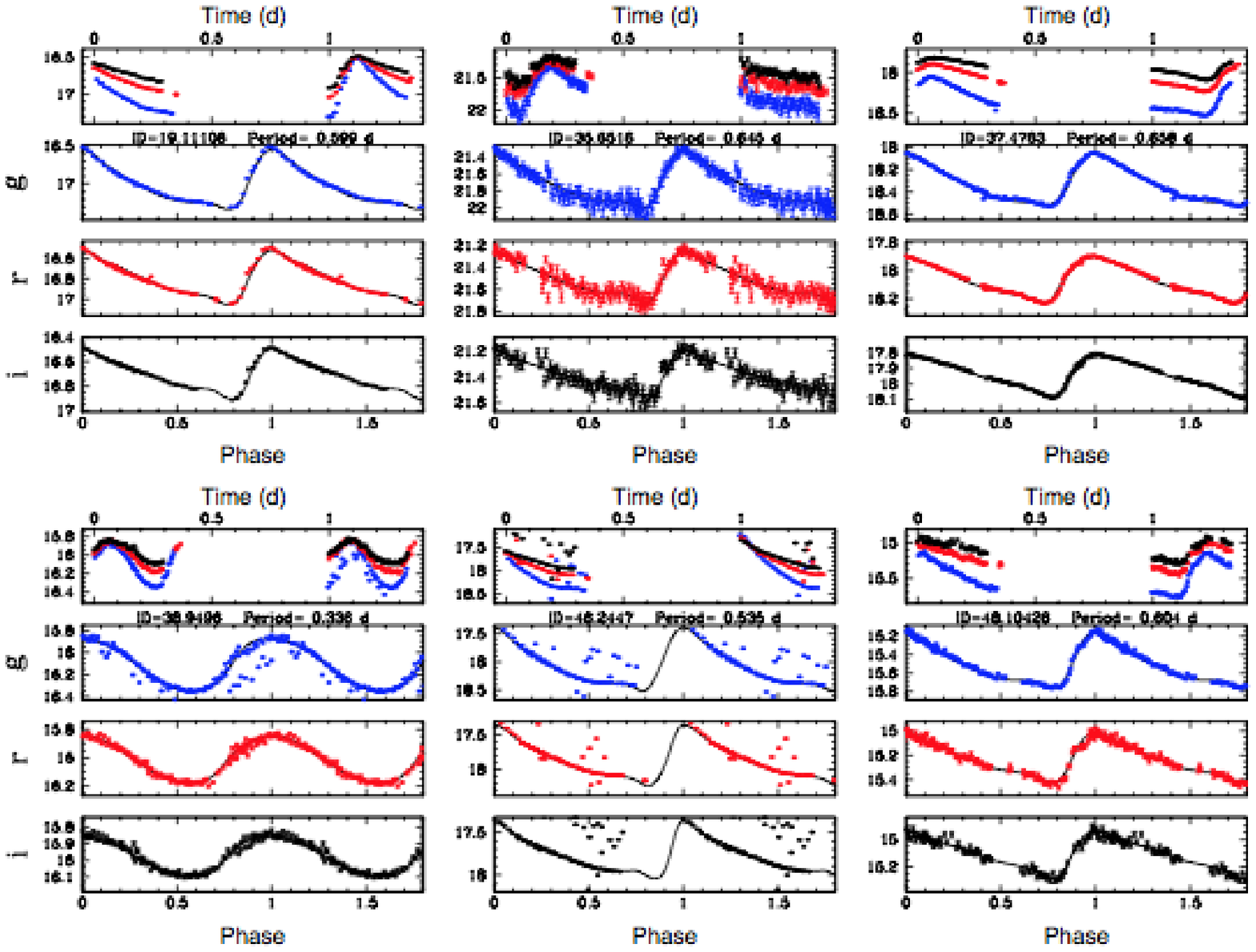}
\caption{Light curves of the six RR Lyrae stars found in the field. The top panel for each star is the 
time series in days from the time of the first observation. The three lower panels show the phased light curve. 
The solid lines are the best template fit to the light curves (see text).
Star 35.6516 (top, middle panel) is likely associated with Hydra II.} 
\label{fig-rr}
\end{figure}

Six RR Lyrae stars were found in the observed field. Five of them are of type {\it ab} and one is of
type {\it c}. Only one of these RR Lyrae stars (ID=35.6516) appears to be associated with Hydra II (see \S~\ref{sec-vH}).
Figure~\ref{fig-rr} shows the light curve of the RR Lyrae stars. We fit light-curve templates to the
data using the set derived by \citet{sesar10} from multi--band light curves of
RR Lyrae stars in SDSS stripe 82. To fit each template we used $\chi^2$ minimization 
\citep[following][]{vivas08}, allowing for
variations around the period given by the LK method, and the amplitude, maximum magnitude and initial phase
derived from the data. 
The mean magnitude of the RR Lyrae 
stars in each band was calculated by transforming the fitted template to intensity, 
integrating under the curve, and transforming the mean back to magnitudes. 
Interstellar extinction values were obtained using the re--calibration made by \citet{schlafly11}
of the \citet{schlegel98} dustmaps.

The mean magnitude of the RR Lyrae stars can be used to derive the distance to the
stars. For this, we used the relationships provided by \citet{caceres08}, which are based on theoretical models
for the SDSS bandpasses.
These models are  consistent with a distance modulus to the Large Magellanic Cloud of $(m-M) _{0}=18.47$ mag
\citep{catelan04}. 
\citet{caceres08} recommend using the longer wavelength 
infrared band for deriving distances because the $g$ and $r$ bands do not show tight period-luminosity relationships
for RR Lyrae stars. We therefore use:

\begin{equation}
M_i = 0.908 - 1.035 \, \log P + 0.220 \, \log Z
\end{equation}

\noindent
which has an uncertainty of 0.045 mag.
The following relationship between $Z$ and [Fe/H] is also taken from \citet{caceres08}:

\begin{equation}
\log{Z} = {\rm [Fe/H]} + \log{(0.638 \times 10^{[\alpha/\rm{Fe}]} + 0.362)}- 1.765
\end{equation}

We assumed [Fe/H]$=-1.65$ as the mean metallicity of the Galactic halo \citep{suntzeff91}, except for star 35.6516,
which is likely to be associated with Hydra II. For this star we used [Fe/H]$=-2.02$, which was directly measured by 
\citet{kirby15} for Hydra II stars. The $\alpha$ abundances ($[\alpha/\rm{Fe}]$) were assumed as 0.2 and 0.3 for the halo and
dwarf galaxies respectively, based on \citet{pritzl05}. The latter is appropriate for dwarf galaxies with [Fe/H]$\sim -2.0$.
To estimate the error we took into account the photometric error 
($\sigma_i \lesssim 0.01$ mags for the 
brighest stars, $\sim 0.05$ mag in for 35.6516), calibration errors ($\sigma_{cal} \sim 0.02, 0.03, 0.02$ in $g$, $r$ and $i$ respectively), error in $E(B-V)$
\citep[10\%,][]{schlegel98}, uncertainty in the $M_i$ relationship ($0.045$ mag), period uncertainty ($\sigma_P = 0.001$ days), 
and abundance dispersion. For the halo stars we assumed 
$\sigma_{\rm [Fe/H]}=0.3$ \citep{suntzeff91}; for Hydra II, \citet{kirby15} measured a dispersion of 0.4 dex. For the $\alpha$
abundances we assumed uncertainties of 0.1 dex. The final error in the true distance
 modulus($\mu_0$) was calculated using:

\begin{equation}
\sigma_{\mu_0}^2 = \sigma_i^2 + \sigma_{cal}^2 + \sigma_{A_i}^2 + \sigma_{M_i}^2 
\end{equation}

\noindent
where

\begin{equation}
\sigma_{M_i}^2 = (0.045)^2 + \left (\frac{1.035}{P \, \ln 10}\right )^2 \sigma_P^2 + (0.220)^2 \sigma_{\rm [Fe/H]}^2 + \left ( \frac{0.220 \times 0.638 \times 10^{[\alpha/\rm{Fe}]} }{0.638 \times 10^{[\alpha/\rm{Fe}]} +0.362} \right )^2 \sigma_{[\alpha/\rm{Fe}]}^2
\end{equation}

As a double check, we calculated the distance in an independent way by transforming the $g,r$ magnitudes to $V$ using 
the transformation equations derived by R. Lupton and available on the SDSS 
webpage\footnote{\url http://classic.sdss.org/dr4/algorithms/sdssUBVRITransform.html\#Lupton2005},

\begin{equation}
V=g - 0.5784 \, (g-r) - 0.0038,
\label{eq-transf}
\end{equation}

\noindent
and then applying the
$M_V$-[Fe/H] relationship given in \citet{cacciari03},

\begin{equation}
M_V  = (0.23 \pm 0.04){\rm [F e/H ]} + (0.93 \pm 0.12)
\end{equation}

The results from both methods are compatible within their errors and are given in Table~\ref{tab-RR}.
The RR Lyrae star associated with Hydra II is highlighted in boldface.
We adopted the distance derived from the $i$ magnitudes as they are more precise than the ones derived using the $V$ magnitudes.

\subsection{Eclipsing Binaries}

\begin{figure*}
\center
\includegraphics[scale=0.7]{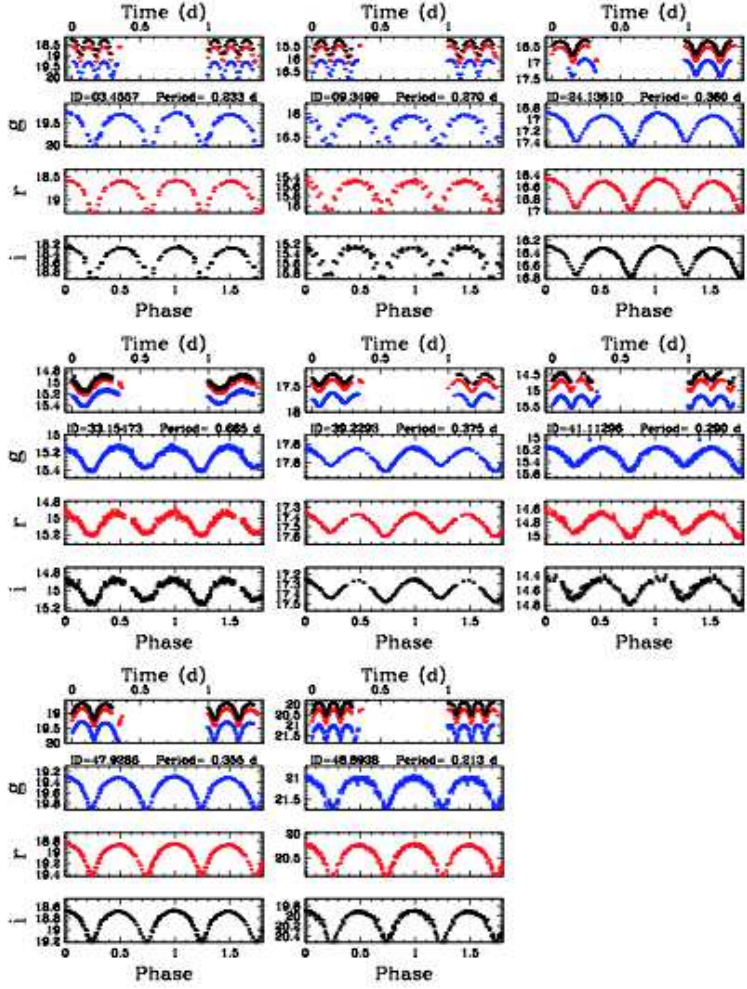}
\caption{Light curves of eight contact eclipsing binaries (EW) with eclipses of similar depths. Format 
is similar to that in Figure 4.}
\label{fig-WUMa}
\end{figure*}

\begin{figure*}
\center
\includegraphics[scale=0.5]{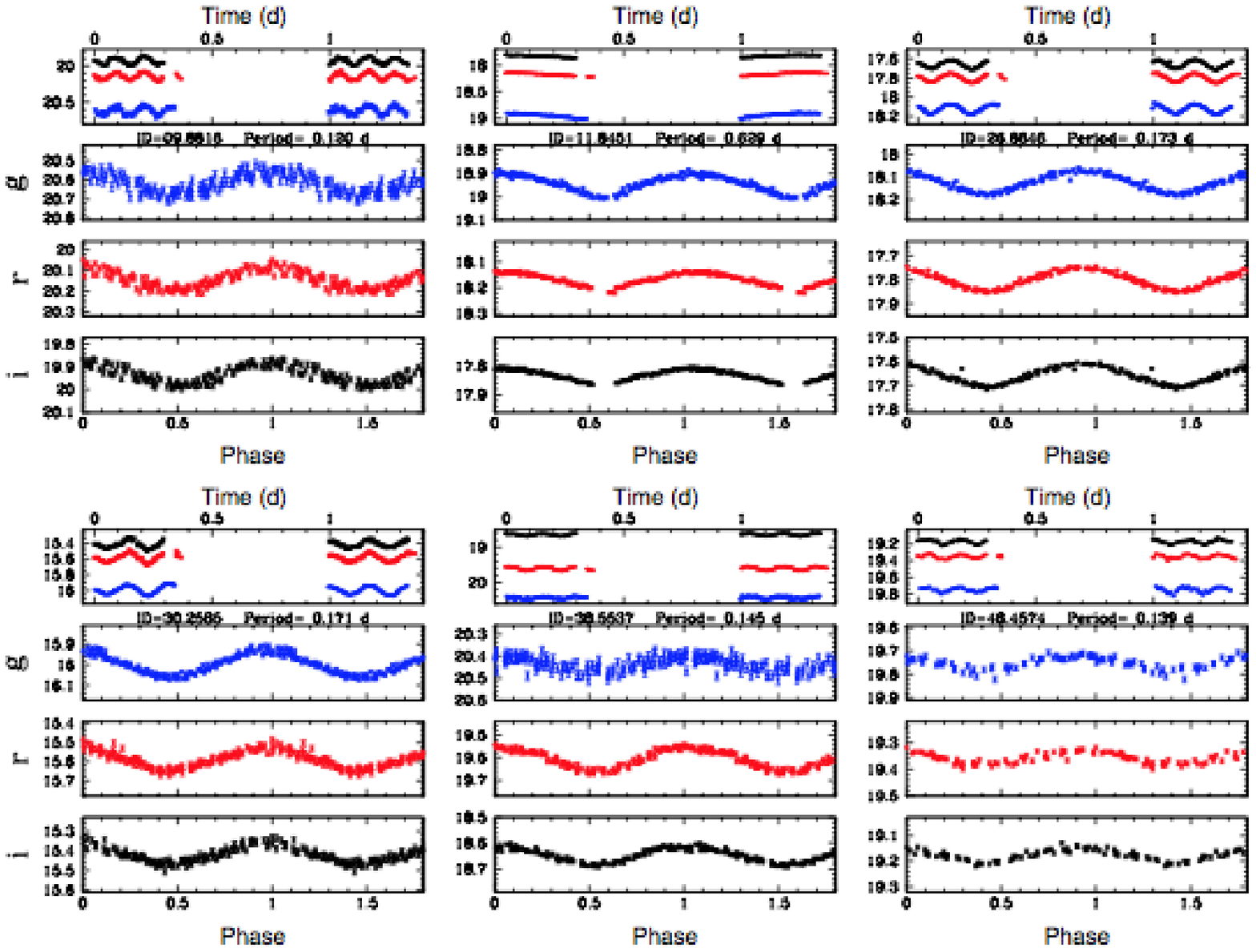}
\caption{Light curves of six contact binaries (EW) with sinusoidal light curves. The format 
is similar to that in Figure 4.}
\label{fig-sin}
\end{figure*}

\begin{figure*}
\center
\includegraphics[scale=0.5]{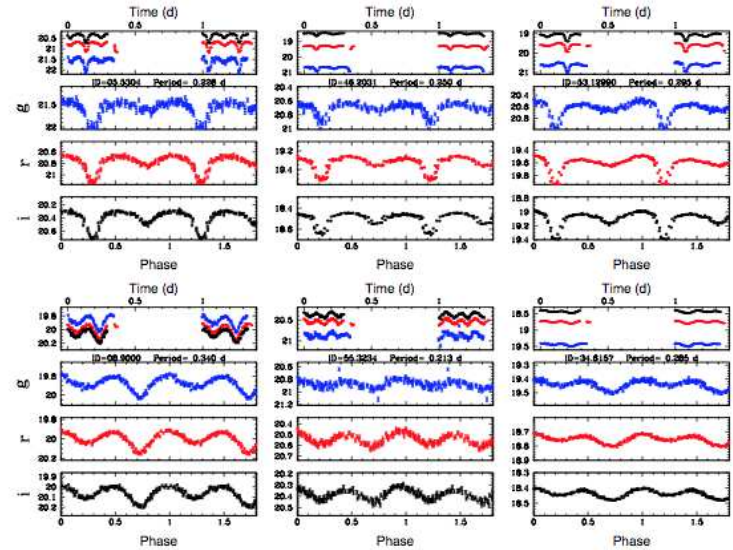}
\caption{Light curves of the six semi-detached (EB) eclispsing binaries. The depth of the eclipses is quite
different and the  stars are distorted into ellipsoids. Format 
is similar to that in Figure 4.}
\label{fig-EB}
\end{figure*}

Twenty-two eclipsing binaries were found in the field. 
Figures~\ref{fig-WUMa} -~\ref{fig-DCEA} show their light curves, which have been grouped according to their shape.

In the first group we have eight contact binaries (EW), which have the typical shape for W UMa stars with
rounded tops and eclipses that in most cases show only slight differences in depth (Figures~\ref{fig-WUMa}).
All of these stars have periods between 0.2 and 0.6 d. 
Then, we have a group of six stars with sinusoidal light curves (Figure~\ref{fig-sin}). 
These stars are also contact eclipsing binaries of the type EW.
Because of their light-curve shape, they are sometimes confused with {\rrc } \citep[see discussion by][]{drake14}. 
We do not believe these stars are {\rrc } because their periods are
too short and their colors are red in all cases except one (see below). 
If the stars in this group are true EW type, then it is very likely that their periods are double those found by the LK algorithm (the LK algorithm 
value is the one reported in Table~\ref{tab-var}) because a full period would have two eclipses.

The stars in Figure~\ref{fig-EB} are semi-detached binaries (EB or $\beta$ Lyrae), which show significant
variations in the depth of the eclipses. Finally, two detached binaries (EA or Algol type) are shown
in the top row (left and middle panels) of Figure~\ref{fig-DCEA}. These stars show almost flat maxima
and V-shaped eclipses.

A few eclipsing binaries are located in the color range of the instability strip 
(see Figure~\ref{fig-CMDall}).
However, their classification is unmistakable because of the shape of their light curves 
(see for example star 33.15473, which shows eclipses of different depth). Only in one of these cases
(star 26.8646 at $(g,g-r)=(17.9,0.58)$) is the light curve sinusoidal and may resemble that
of type {\it c} RR Lyrae stars. However, its short period (0.17 d) and low amplitude (Amp g $= 0.13$)
make it very unlikely for the star to be a real {\rrc } \citep[see][]{palaversa13,drake14}. The star is also located toward the red 
side of the instability strip, which is not the case for {\rrc } stars.

\subsection{Dwarf Cepheid Stars (and an Anomalous Cepheid?)}

\begin{figure*}
\center
\includegraphics[scale=0.5]{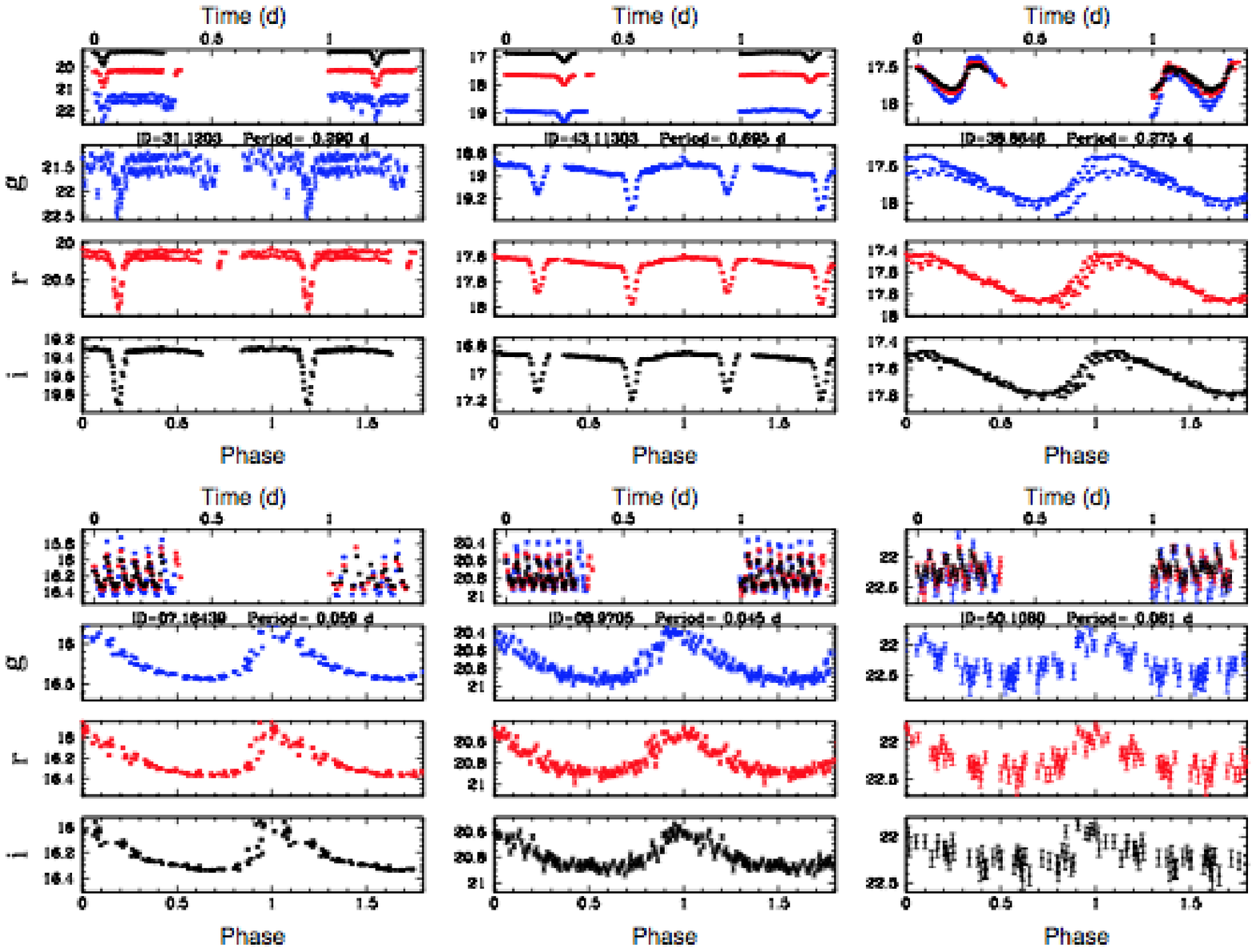}
\caption{Light curves of two detached binaries (top, left and middle) one likely anomalous cepheid (top right)
and three dwarf cepheids (bottom row). Format 
is similar to that in Figure 4.}
\label{fig-DCEA}
\end{figure*}

Dwarf cepheid stars encompass the pulsating variable stars in the region where the instability strip intersects the 
main sequence. Dwarf cepheids from Pop I systems are known as $\delta$ Scuti, while those from Pop II systems are called
SX Phe stars. Not knowing with certainty the stellar population to which the stars in our field belong, we prefer to name them by the collective name {\sl dwarf cepheids} \citep{mateo98}. The main characteristic of this group of stars is their very short period (just a few hours). In dSph galaxies
these stars show large amplitudes \citep[$\Delta V \sim 0.5$ mags,][]{vivas13}.
We have found four stars that have the characteristics of dwarf cepheid stars, namely, 07.16439, 08.9705, 50.1080 and
38.8646 (Figure~\ref{fig-DCEA}). The first three in this group (bottom row in Figure~\ref{fig-DCEA}) have exactly these properties, with
periods ranging between 0.04 and 0.06 d. We cannot a priori calculate the distance to these stars without knowing their pulsation mode. However, applying
the Period-Luminosity relationships by \citet{nemec94} for both fundamental and first overtone modes, we find that stars 7.16439, 8.9705 and 50.1080 
must be located at 3 or 4 kpc, 27 or 31 kpc and
63 or 72 kpc, respectively. To obtain these numbers, we transformed first the mean $g, r$ magnitudes to Johnson $V$ (see above). 
Thus, independent of their pulsation mode, they should be halo field stars and the more likely correct classification for these stars is therefore SX Phe. Their large amplitude supports this classification
\citep{breger00}.
The last star in this group (38.8646, top/right in Figure~\ref{fig-DCEA}) is more difficult to classify.
Its period (0.275 d) is somewhat larger than typical for dwarf cepheid stars. On the other hand, although the 
period and amplitude is right where {\rrc } stars are expected, the shape of the light curve is too 
asymmetric for this class of stars. It is possible that 38.8646 is then an anomalous cepheid. 
According to \citet{pritzl02}, the period for this star may be on the short side for anomalous cepheids (although 
there are stars with periods as short as 0.3 d), but still, considering all of its properties (color, amplitude, 
period, shape), this is the most likely classification. It is possible that this star is displaying a slight change of period from one night to the other since in the first night it shows a full cycle while in the second night, which spans the same time range, the cycle is not complete. This is causing the double sequence in the phased light curves in Figure~\ref{fig-DCEA}.

\section{Variable stars in Hydra II: the information provided by its RR Lyrae star \label{sec-vH}}

\begin{figure*}
\epsscale{0.7}
\plotone{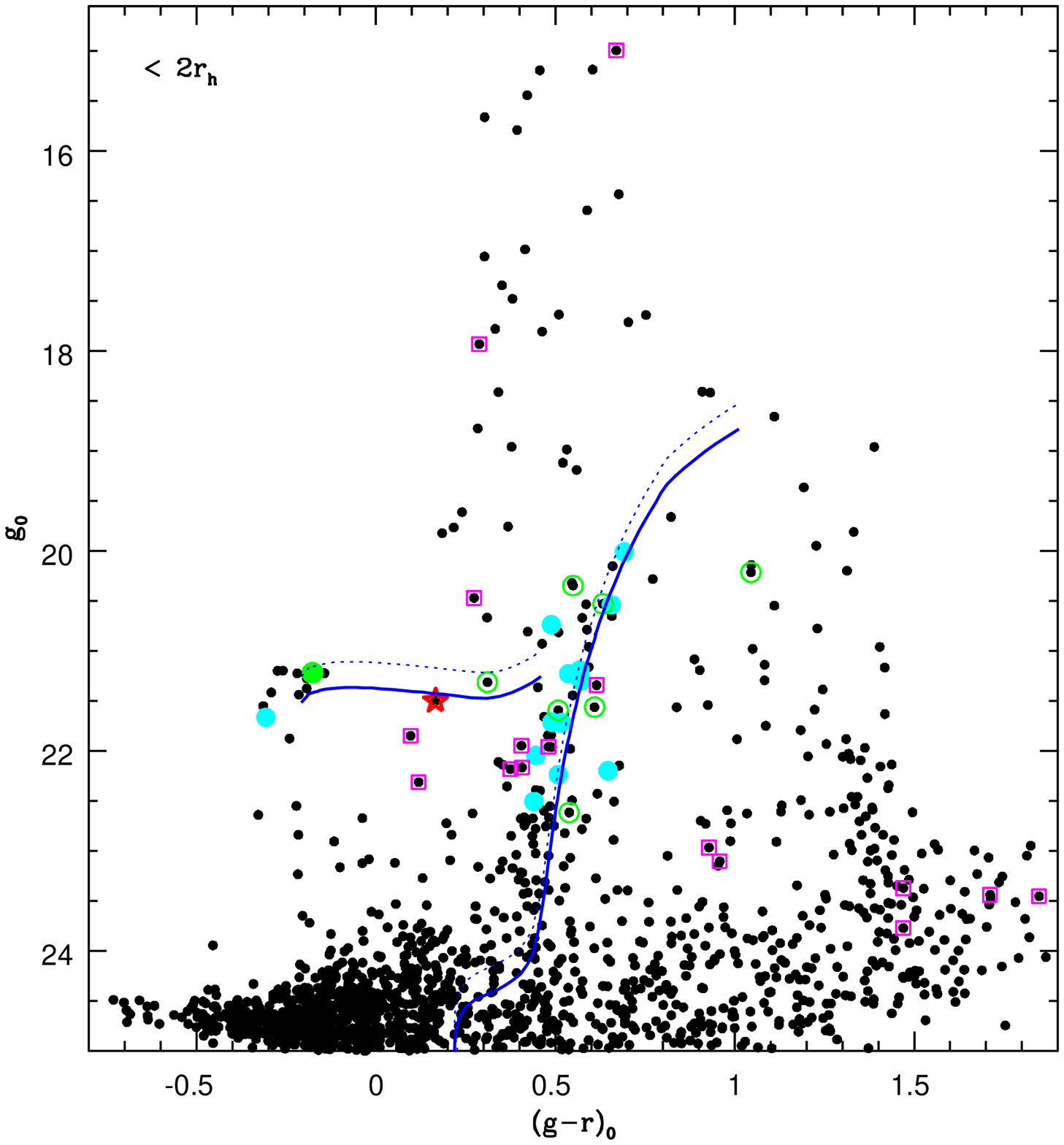}
\caption{Color-Magnitude diagram of stars within $2r_h$ of the center of Hydra II (1462 stars).
The red star indicates the location of the RR Lyrae star 35.6516. For reference we show the stars
confirmed with radial velocities as members of Hydra II \citep{kirby15} as large cyan circles, while
radial velocity non-members are indicated with green circles.
The open magenta squares show the location of variable stars that could not be classified.
Solid and dotted lines correspond to a 13 Gyr old, metal-poor ([Fe/H]$=-2.0$) isochrone 
\citep{bressan12} shifted to 151 and 134 kpc, respectively.}
\label{fig-CMDHyaII}
\end{figure*}

Figure~\ref{fig-CMDHyaII} shows the CMD containing all stars within 2 half-light radii of the galaxy.
The features of the galaxy are visible in this CMD with the top of the red giant branch at $g_0\sim 19.0$, a group of blue horizontal branch (BHB)
stars at $g_0\sim 21.5$  and the main sequence turn-off
at the faint limit of the diagram. As mentioned before,
one RR Lyrae star (35.6516 at $g_0=21.50$) is located well within $1\times r_h$ of the galaxy, at $42\arcsec$ from its center. We have marked this star, at its mean magnitude and 
color, with a red star symbol in 
Figure~\ref{fig-CMDHyaII}. The star lies at the level of the horizontal branch of Hydra II,  hence
 its association with the galaxy
is clear. Unfortunately, although \citet{kirby15} included some potential BHB stars in their spectroscopic sample, star 35.6516 is not among them. 
Thus, no final confirmation from radial velocity is available. 
Needless to say, Galactic halo RR Lyrae stars at distances larger than $\sim 80$ kpc are very rare 
\citep{zinn14,watkins09}. Indeed, notice 
the lack of RR Lyrae stars for $g>18.5$ (Figure~\ref{fig-CMDall}). 
It is highly unlikely that this is 
a field halo star. 

\begin{deluxetable}{lccccccc}
\tabletypesize{\normalsize}
\tablecolumns{8}
\tablewidth{0pc}
\tablecaption{Suspected Variable Stars within $2r_h$ of Hydra II}
\tablehead{
ID & RA (deg) & DEC (deg) & $g_0$ & $(g-r)_0$ & Amp$_g$ & Amp$_r$ & Amp$_i$ \\
}
\startdata
   35.6652 & 185.430960 & -32.037326 & 14.99 & 0.669 &  0.3 &  0.5 &  0.2 \\
   35.7152 & 185.451049 & -32.026565 & 17.93 & 0.288 &  0.2 &  0.1 &  0.1 \\
   35.6106 & 185.422854 & -31.942608 & 20.47 & 0.274 &  0.6 &  0.5 &  0.6 \\
  35.17189 & 185.451179 & -32.026491 & 21.34 & 0.615 &  2.6 &  1.8 &  1.3 \\
  35.17407 & 185.457215 & -31.954393 & 21.85 & 0.097 &  1.0 &  0.8 &  0.5 \\
  35.16949 & 185.441125 & -32.012642 & 21.95 & 0.405 &  0.8 &  0.6 &  0.5 \\
   35.7151 & 185.455066 & -31.946172 & 21.96 & 0.481 &  2.1 &  1.5 &  2.2 \\
  35.15898 & 185.405407 & -32.004393 & 22.17 & 0.407 &  1.4 &  1.4 &  1.1 \\
   35.5412 & 185.405288 & -32.004220 & 22.18 & 0.375 &  1.8 &  0.6 &  0.6 \\
   35.7273 & 185.457382 & -31.954359 & 22.31 & 0.119 &  1.9 &  1.6 &  1.0 \\
\enddata
\label{tab-suspected}
\end{deluxetable}

From Table~\ref{tab-RR} we can put the distance to Hydra II at $151\pm 8$ kpc from the Sun, or
148 kpc from the Galactic center (assuming $R_\odot=8$ kpc). This is $13\%$ further away than the original estimate of the heliocentric distance 
made by \citet[][$134 \pm 10$ kpc]{martin15} based on the average magnitude of several
stars at the horizontal branch. In \citet{martin15}, the RR Lyrae star was measured with $g_0=21.16$ which 
is near maximum light. This surely influenced their distance estimate.
The solid line in Figure~\ref{fig-CMDHyaII} represents an old (13 Gyr) and metal-poor 
([Fe/H]$=-2.0$) Parsec isochrone\footnote{\url http://stev.oapd.inaf.it/cgi-bin/cmd} \citep{bressan12} shifted to a distance of 151 kpc. For reference
we also show the same isochrone at the distance reported in the discovery paper (134 kpc, dotted line).
Radial velocity members and non-members from \citet{kirby15} are shown as cyan and green circles, respectively.
A special case among the non-members is indicated as a solid green circle. Although this star had the correct radial velocity to be a member of Hydra II, it was 
discarded as a member by \citet{kirby15} since it was too bright in their photometry for being an HB star\footnote{The star has ID=191385 in \citet{kirby15} with
$g_0,(g-r)_0 = 20.7, 0.16$ in their photometry.}. Our photometry locates this star among the HB candidates and we conclude it must be a member. Although we do not have an explanation for the
discrepancy, we noticed that the spectra of this star in \citeauthor{kirby15} has a S/N more similar to
stars of magnitude $g_0=21.2$ (our magnitude) than to others stars with $g_0\sim 20.7$ observed with the same slitmask (hence, under identical observing conditions).
The other radial velocity confirmed member in the HB of Hydra II is the faintest object within the group of candidates, at $g_0=21.66$.
There seems to be an apparent inconsistency between the distance given by the RR Lyrae star and the location of the other BHB candidates (at $(g-r)_0\sim -0.25$)
since the latter appear slightly brighter than the isochrone at the HB level. 
Confirmation of the rest of the BHB candidates as members would be needed to resolve this apparent mismatch. But even if all those BHB candidates are real members, there are other possible
explanations for the apparent discrepancy. 
To estimate the distance we used the mean metallicity provided by \citet{kirby15}. However, the
5 stars measured by \citeauthor{kirby15} present a large dispersion in the range 
$-2.76 \leq {\rm [Fe/H]} \leq -1.89$. Using the highest metallicity measured by \citeauthor{kirby15}
(which is a perfectly reasonable value for an RR Lyrae star), we obtained a heliocentric distance of
149 kpc. Thus, the discrepancy would be shortened (or disappear) if the metallicity of the RR Lyrae
star is higher than the mean of the galaxy. We note that a dispersion in metallicity of 0.4 dex was included in the
error propagation of the distance modulus, and thus, this lower value of distance is within our error
estimation. Another possibility for the apparent discrepancy is that the RR Lyrae stars and the BHB stars 
have different evolutionary stage, with the latter being more evolved than the RR Lyrae star.

The increase in heliocentric distance also means that the physical size is larger and the galaxy is more luminous than previously thought. 
The revised values for these quantities are $r_h = 76^{+12}_{-10}$ pc and
$M_V = -5.1 \pm  0.3$, which re-affirms the nature of this system as a dwarf galaxy since it further separates it from the more compact group of globular clusters.

The possible association of Hydra II with the Magellanic Clouds \citep{martin15} is still
compatible with the new distance derived here. 
Hydra II still appears to lead the Clouds, being at a distance and location in the sky that is consistent with the future orbit of the LMC.  It resides $\sim 23$ kpc from the LMC's orbital plane. 
Further indications of this association come from the spectroscopic work by \citet{kirby15}, who measured a radial velocity similar to the gas in the Magellanic stream at that position in the sky.

We note that there are no more RR Lyrae stars of similar magnitude in the neighborhood that could be interpreted as possible tidal debris material.  

The lack of anomalous cepheids in Hydra II indicates that younger stellar populations may not exist in this galaxy or, if they exist, that
they are not significant. On the other hand SX Phe stars are expected at 1.2 to 3 mag below the
horizontal branch \citep{vivas13}. Thus, if Hydra II has these type of stars they should have magnitudes between 22.9 and 24.7. Our photometric errors
increase significantly in this magnitude range ($\sigma>0.2$ mags), potentially explaining why we may have not detected SX Phe stars in the galaxy. The faintest star 
we detected in the field is an SX Phe at $g=22.3$ but it is located too far from the center of Hydra II 
for us to suspect any relationship between them.

Figure~\ref{fig-CMDHyaII} also shows the location of other variable stars that do not show
periodic variability on scales of $<1$ day. We examined the time series for each one of these stars 
including data from the discovery paper \citep{martin15}, which date from March 2013 ($\sim 700$ days 
before the dataset presented here). No firm conclusions were drawn from this exercise, although some
of these stars may be long period variables. Except for the stars at the faint end ($g>23$), all others seem to be genuine variables. The position and observed amplitudes for these suspected variables (10 stars) are reported in Table~\ref{tab-suspected}.

\section{RR Lyrae stars in ultra--faint dwarf galaxies}

All dwarf galaxies that have been searched for variable stars, including the ultra--faint galaxies, have at least 
one RR Lyrae star. 
In Table~\ref{tab-NRR} we show the number of RR Lyrae stars in all of the ultra faint dwarfs that have been explored
to date, sorted by decreasing number of $M_V$.  Data for the absolute magnitudes and the number of RR Lyrae stars for most systems were taken from \citet{sand12} 
and \citet{baker15}, respectively. Data for Bootes III were taken from \citet{sesar14}
and \citet{correnti09}, and for Hydra II we adopted the values derived in this paper.
Of all the galaxies with only one RR Lyrae star, Bootes III and Hydra II are the most luminous. 
There are several galaxies which are less luminous than Hydra II and contain one, or even two, RR 
Lyrae stars.

\begin{deluxetable}{lcrc}
\tabletypesize{\normalsize}
\tablecolumns{4}
\tablewidth{0pc}
\tablecaption{Number of RR Lyrae Stars in Ultra-Faint Dwarfs of the Milky Way}
\tablehead{
Galaxy & $M_V$  &  $N_{RR}$ \\
}
\startdata
Bootes I              & $-6.3\pm0.2$ & 15 \\
Hercules             & $-6.2\pm0.4$ &  9 \\
Bootes III            & $-5.8\pm0.5$ &  1  \\
Ursa Major I        & $-5.5\pm0.3$ &  7 \\
Leo IV                 & $-5.5\pm0.3$ &  3 \\
Hydra II              & $-5.1\pm0.3$ &  1  \\
Canes Venatici II & $-4.6\pm0.2$ &  2 \\
Ursa Major II       & $-4.0\pm0.6$ &  1  \\
Coma Berenice   & $-3.8\pm0.6$ &   2 \\
Segue2               & $-2.5\pm0.2$ &  1 \\
BootesII              & $-2.2\pm0.7$ &  1 \\
Segue1               & $-1.5\pm0.5$ &  1 \\
\enddata
\label{tab-NRR}
\end{deluxetable}

\begin{figure*}
\epsscale{0.7}
\plotone{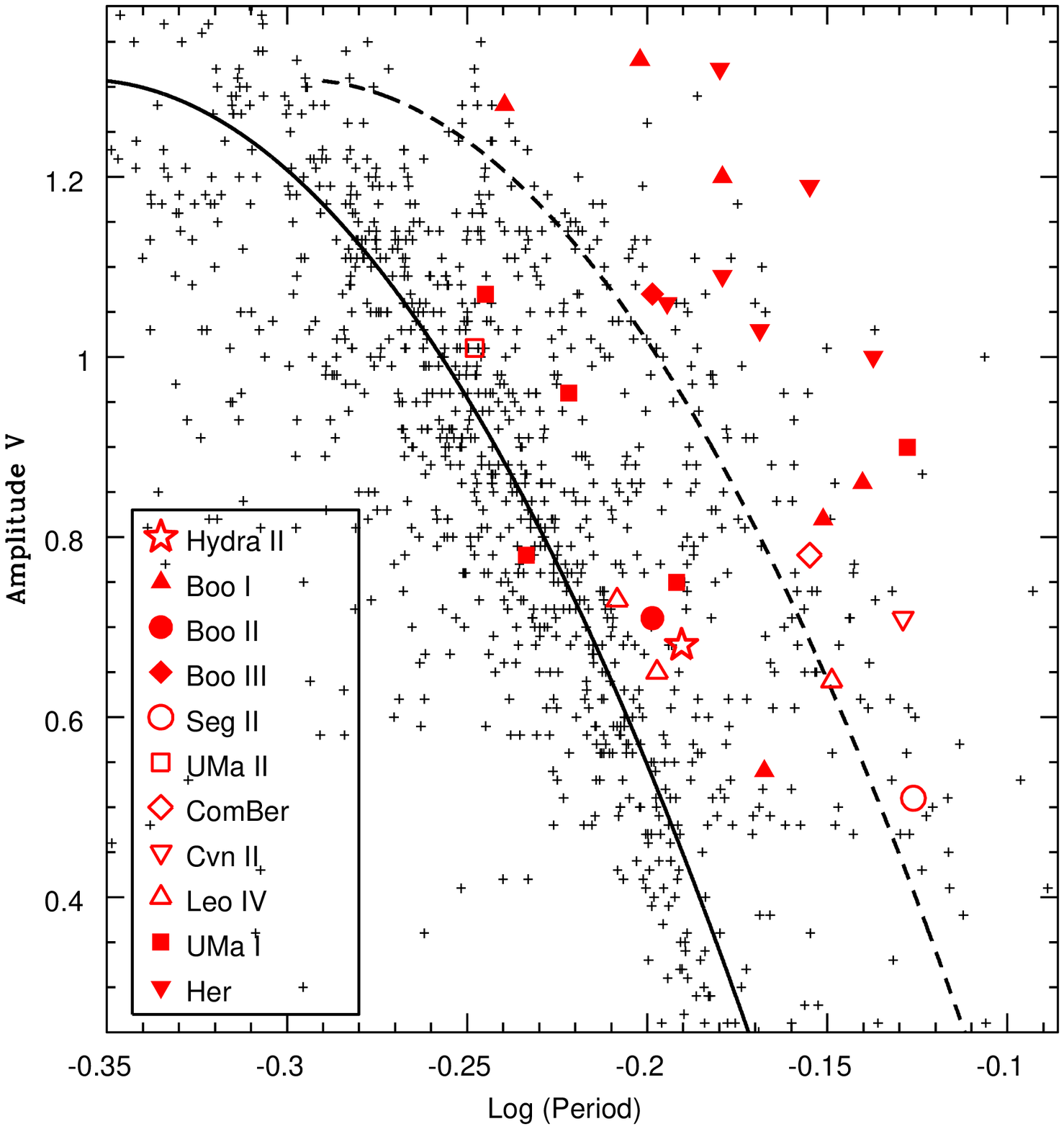}
\caption{Period--Amplitude diagram for {\rrab } stars. 
Crosses represent the Milky Way halo
field population from the La Silla-QUEST survey \citep{zinn14}. 
The solid and dashed lines represent the Oo I and Oo II locus, respectively, as defined by \citet{zorotovic10}. Data for RR Lyrae stars in the ultra-faint galaxies come from \citet[][Bootes I]{siegel06},
\citet[][Hercules]{musella12}, \citet[][Ursa Major I]{garofalo13}, \citet[][Leo IV]{moretti09}, \citet[][Canes Venatici II]{greco08}, \citet[][Coma Berenices]{musella09},
\citet[][Segue II]{boettcher13}; RR Lyrae stars in Bootes II and Bootes III, and in Ursa Major II were taken from \citet{sesar14} and \citet{dallora12} respectively
but their light-curve parameters were re--calculated as explained in the main text.
}
\label{fig-Oo}
\end{figure*}

\bigskip

Globular clusters with 5 or more RR Lyrae stars  are classified in Oosterhoff groups I and II based on the mean periods of their {\rrab } stars;
the mean periods are 0.55 and 0.65 days for Oo I and OoII systems, respectively \citep{catelan15}. 
The mean periods of {\rrab } in a system or, more generally, the distribution of the RR Lyrae stars in the Period-Amplitude plane (or Bailey diagram) 
and its comparison with the locus in the same plane of RR Lyrae stars in Oo I and Oo II clusters, is also often used as reference to classify a stellar system in Oo I or Oo II like.  
Figure~\ref{fig-Oo} shows the location of the Hydra II RR Lyrae star in the Bayley diagram, together with the location of the {\it ab}-type RR Lyrae in the other ultra--faint galaxies ($M_V>-7$) that have been explored so far
for variable stars (11 galaxies including Hydra II)\footnote{We are not including Segue 1 because although \citet{simon11} reports one RR Lyrae variable in this faint system
based on spectroscopic variability, no accurate period or amplitude of the light curve has been derived for it.}.
To include the RR Lyrae variable of Hydra II in this diagram, we transformed first the best fit
light curves templates in $g$ and $r$ into a $V$ light-curve using Equation~\ref{eq-transf}. Our estimate for the $V$ amplitude of star 35.6516 is 0.56 mags.
For Bootes I, only $B$ band amplitudes are available \citep{siegel06}. We transformed to $V$ amplitudes using 
$\Delta V = 0.790 \Delta B - 0.006$, which was obtained from combining the $\Delta V-\Delta I$ and $\Delta B-\Delta I$ equations in \citet{dorfi99}.
In the case of Bootes II, Bootes III and Ursa Major II, we revised the light curve parameters as explained in the Appendix. 

Based on the period of the only RR Lyrae star in this system,
Hydra II is then classified as an Oo II system. However, based on the Period-Amplitude diagram, 
Hydra II could be classified as an Oo-intermediate galaxy. This ambiguity in the classification is a 
reminder that Oosterhoff groups are based on mean properties of systems with several RR Lyrae stars. Most clusters present a spread around the mean values of the periods or around the loci in Bailey's diagram \citep[see discussion in][]{catelan09,catelan15} and so, one star alone
may not indicate the right group. In Hydra II, as it is the case for several other of the Milky Way's ultra--faint dwarfs, only one {\rrab } is present. 

We also show in Figure~\ref{fig-Oo} the distribution of halo field stars
from the La Silla--QUEST survey \citep{zinn14}.
It is clear that while the halo stars concentrate toward the Oo I locus, most of the RR Lyrae stars in the ultra-faint dwarfs (including Hydra II) are mainly located in the 
region of the Oo-intermediate and OoII loci.
The mean period of the ensemble of 28 RR Lyrae stars in 11 ultra--faint dwarfs is 
0.667 d. This value is significantly different from the mean value of the RR Lyrae stars in the halo (0.586 d) and in large satellite galaxies like the LMC (0.576 d), SMC (0.596 d),
Sagittarius (0.575 d) and Fornax (0.585 d) \citep{zinn14}. 
If the Milky Way was formed by the accretion of small sub-structures as expected by the 
LCDM cosmology,
the ultra--faint galaxies 
may contribute to the Oo II tail of the population of halo RR Lyrae, but they do not seem 
to be the source of the main population. As suggested by \citet{zinn14} 
the accretion of large galaxies similar to the LMC, SMC, Sgr or Fornax seems to be 
needed to assemble the RR Lyrae population of the halo. \citet{fiorentino15} proposed a
similar scenario based on the existence of high amplitude short period (HASP) RR Lyrae stars in
the Magellanic Clouds and Sgr dSph. They discarded Fornax-like galaxies as progenitor based on the lack of HASP stars in that galaxy. Similarly, as shown in Figure~\ref{fig-Oo}, 
not one of the RR Lyrae stars in the ultra--faint dwarf
galaxies are located within the HASP region.  

\section{Conclusions}

We obtained high--cadence time series photometry in the $g$, $r$, and $i$ bands in the field of the 
newly discovered ultra--faint galaxy Hydra II with DECam at CTIO to search for and characterize 
short period variable stars. We found 32 periodic variable stars in the field but only one, an RR Lyrae star, is located within the half--light radius of Hydra II. Assumed to be a Hydra II member, this star allowed us to
improve the distance estimation to the galaxy. The distance calculation yielded 151 kpc from the Sun, or 148 kpc from the Galactic center, a distance which is $\sim 13\%$ farther away
than the original estimate in the discovery paper \citep{martin15}. The larger distance also implies a larger physical size, $r_h = 76^{+12}_{-10}$ pc, and brighter absolute magnitude,
$M_V = -5.1 \pm  0.3$, for Hydra II.

No anomalous cepheids were found in the galaxy, suggesting that Hydra II does not contain a significant intermediate--age population.

The pulsational properties of the Hydra II RR Lyrae star are similar to those in other ultra-faint galaxies and they are different to the bulk of RR Lyrae stars in the halo of the Milky Way.
Thus, in a hierarchical formation scenario, ultra--faint galaxies may have contributed to the long-period tail (Oo II group) of the period distribution of field RR Lyrae stars but accretion of large galaxies such as the Magellanic Clouds, Sagittarius or Fornax is needed to account for the predominantly Oo I population in the galactic halo.

Until now, all dwarf galaxies that have been adequately searched for variable stars have yielded RR Lyrae stars, which confirms that these systems predominantly contain old,
metal-poor stellar populations. Hydra II is no exception.

\acknowledgments

Based on observations at Cerro Tololo Inter-American Observatory, National Optical Astronomy Observatory (NOAO Prop. ID: 2013B-0440; PI: Nidever), which is operated by the Association of Universities for Research in Astronomy (AURA) under a cooperative agreement with the National Science Foundation. 
This project used data obtained with the Dark Energy Camera (DECam), which was constructed by the Dark Energy Survey (DES) collaboration.
Funding for the DES Projects has been provided by 
the U.S. Department of Energy, 
the U.S. National Science Foundation, 
the Ministry of Science and Education of Spain, 
the Science and Technology Facilities Council of the United Kingdom, 
the Higher Education Funding Council for England, 
the National Center for Supercomputing Applications at the University of Illinois at Urbana-Champaign, 
the Kavli Institute of Cosmological Physics at the University of Chicago, 
the Center for Cosmology and Astro-Particle Physics at the Ohio State University, 
the Mitchell Institute for Fundamental Physics and Astronomy at Texas A\&M University, 
Financiadora de Estudos e Projetos, Funda{\c c}{\~a}o Carlos Chagas Filho de Amparo {\`a} Pesquisa do Estado do Rio de Janeiro, 
Conselho Nacional de Desenvolvimento Cient{\'i}fico e Tecnol{\'o}gico and the Minist{\'e}rio da Ci{\^e}ncia, Tecnologia e Inovac{\~a}o, 
the Deutsche Forschungsgemeinschaft, 
and the Collaborating Institutions in the Dark Energy Survey. 
The Collaborating Institutions are 
Argonne National Laboratory, 
the University of California at Santa Cruz, 
the University of Cambridge, 
Centro de Investigaciones En{\'e}rgeticas, Medioambientales y Tecnol{\'o}gicas-Madrid, 
the University of Chicago, 
University College London, 
the DES-Brazil Consortium, 
the University of Edinburgh, 
the Eidgen{\"o}ssische Technische Hoch\-schule (ETH) Z{\"u}rich, 
Fermi National Accelerator Laboratory, 
the University of Illinois at Urbana-Champaign, 
the Institut de Ci{\`e}ncies de l'Espai (IEEC/CSIC), 
the Institut de F{\'i}sica d'Altes Energies, 
Lawrence Berkeley National Laboratory, 
the Ludwig-Maximilians Universit{\"a}t M{\"u}nchen and the associated Excellence Cluster Universe, 
the University of Michigan, 
{the} National Optical Astronomy Observatory, 
the University of Nottingham, 
the Ohio State University, 
the University of Pennsylvania, 
the University of Portsmouth, 
SLAC National Accelerator Laboratory, 
Stanford University, 
the University of Sussex, 
and Texas A\&M University.
We thank the anonymous referee for interesting suggestions that improved the content of this paper.
SM acknowledges support from grant NSF-AST1312863.

{\it Facility:} \facility{Blanco (DECam)}

\appendix

\section{Revision of light curve parameters of RR Lyrae stars in Bootes II, Bootes III and Ursa Major II}

One RR Lyrae star of the type {\it ab} has been identified in each of the ultra-faint galaxies Bootes II,
Bootes III and Ursa Major II. Data for these three stars are available on the CRTS webpage and with a large 
number of epochs ($\sim 100-300$) to ensure good light curve parameters. 
\citet{dallora12} presented a small number of observations for the RR Lyrae star in Ursa Major II and 
they warned about a possible different period. This star has 136 epochs in CRTS and we found 
a period of 0.56512 d. No periodic signal near the 0.6593 d given by \citet{dallora12} was found with the CRTS data.
For the RR Lyrae star in Bootes III, \citet{sesar14} report multi-epoch observations in the Mould--R
band from the Palomar Transient Factory (PTF). We used the CRTS data to derive the 
parameters for the $V$ band and thus were able to directly compare with other RR Lyrae stars in ultra--faint dwarfs in the Period--Amplitude diagram. 
Finally, \citet{sesar14} also report one RR Lyrae in Bootes I from the CRTS.
However, the authors fitted the light curve using templates in the $r$ band. Here we use $V-$band templates to re--derive the parameters of its lightcurve. The results we obtained for this star are quite similar to those by \citet{sesar14}.

\begin{figure*}
\epsscale{0.7}
\plotone{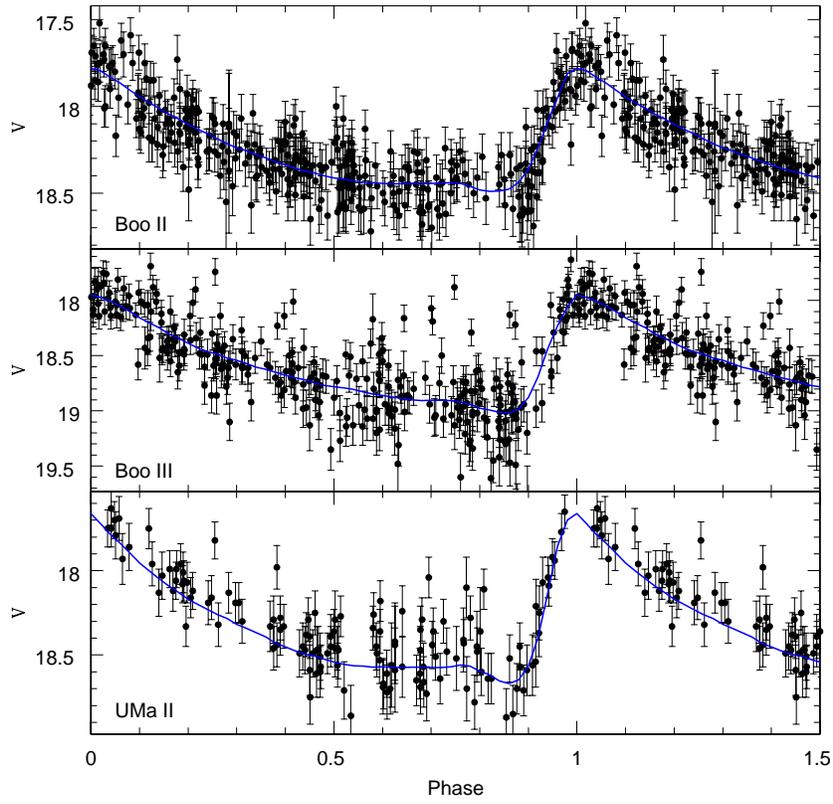}
\caption{Light curve for RR Lyrae stars in Bootes II, Bootes III and Ursa Major II from the CRTS. Solid lines represent the best fit template to the light curves.}
\label{fig-od}
\end{figure*}

We downloaded the light curve for the three variable stars in the 
CRTS\footnote{\url http://nesssi.cacr.caltech.edu/DataRelease/} Data Release 2 \citep[CRTS,][]{drake09} based on the coordinates provided by
\citet{sesar14} and \citet{dallora12}. We fit templates to the light curves following the same procedure described above for Hydra II. In this case however, we 
used the templates provided by \citet{layden98} because they are more appropriate for $V-$band observations. The re--derived light curves for the stars in these three galaxies are shown in Figure~\ref{fig-od} and the parameters are in Table~\ref{tab-od}. The table includes the period found previously by \citet{dallora12} and \citet{sesar14}.

\begin{deluxetable}{lccccccl}
\tabletypesize{\footnotesize}
\rotate
\tablecolumns{8}
\tablewidth{0pc}
\tablecaption{Light curve parameters for the RR Lyrae stars in Bootes II, Bootes III and Ursa Major II}
\tablehead{
Galaxy & ID (CRST)  &  Period (d) & Amp (V) & N & $\langle V \rangle$ & Previous Period (d) & Previous Reference \\
}
\startdata
Bootes II & CSS\_J135807.0+125123 &  0.66349 & 0.71 & 349 & 18.23 & 0.63328 & \citet{sesar14} \\
Bootes III & CSS\_J140034.5+255552 & 0.63328 & 1.07 & 310 & 18.56 & 0.63327 & \citet{sesar14} \\
Ursa Major II & CSS\_J085037.5+631009 & 0.56512 & 1.01 & 136 & 18.24 &  0.6593 & \citet{dallora12} \\
\enddata
\label{tab-od}
\end{deluxetable}

\end{document}